\documentclass{article}

\PassOptionsToPackage{round}{natbib}

 \usepackage[preprint]{nips_2018}

\usepackage[utf8]{inputenc} 
\usepackage[T1]{fontenc}    
\usepackage{hyperref}       
\usepackage{url}            
\usepackage{booktabs}       
\usepackage{amsfonts}       
\usepackage{nicefrac}       
\usepackage{microtype}      
\usepackage{graphicx}
\usepackage{algorithm}
\usepackage{algorithmic}
\usepackage{amsmath,amssymb,amsfonts,mathrsfs}
\usepackage{multicol}
\usepackage{subfig}
\usepackage{xcolor}

\newcommand{\R}{\mathbb{R}}

\title{Inference of the three-dimensional chromatin structure and its temporal behavior}

\author{
  Bianca-Cristina Cristescu \\
  ETH Zurich, IBM Research Zurich \\
  \texttt{cristesc@ethz.ch} \\
  \And
  Zalán Borsos \\
  ETH Zurich \\
  \texttt{zalan.borsos@inf.ethz.ch} \\
  \And
  John Lygeros \\
  ETH Zurich \\
  \texttt{jlygeros@ethz.ch} \\
  \And
 Mar\'ia Rodr\'iguez Mart\'inez \\
  IBM Research Zurich \\
  \texttt{mrm@zurich.ibm.com} \\
  \And
  Maria Anna Rapsomaniki \\
  IBM Research Zurich \\
  \texttt{aap@zurich.ibm.com} \\
}

\begin{document}

\maketitle

\begin{abstract}
Understanding the three-dimensional (3D) structure of the genome is essential for elucidating vital biological processes and their links to human disease. To determine how the genome folds within the nucleus, chromosome conformation capture methods such as HiC have recently been employed. However, computational methods that exploit the resulting high-throughput, high-resolution data are still suffering from important limitations. In this work, we explore the idea of manifold learning for the 3D chromatin structure inference and present a novel method, \emph{REcurrent Autoencoders for CHromatin 3D structure prediction (REACH-3D)}. Our framework employs autoencoders with recurrent neural units to reconstruct the chromatin structure. In comparison to existing methods, REACH-3D makes no transfer function assumption and permits dynamic analysis. Evaluating REACH-3D on synthetic data indicated high agreement with the ground truth. When tested on real experimental HiC data, REACH-3D recovered most faithfully the expected biological properties and obtained the highest correlation coefficient with microscopy measurements. Last, REACH-3D was applied to dynamic HiC data, where it successfully modeled chromatin conformation during the cell cycle.
\end{abstract}

\section{Introduction}
In eukaryotic cells, the total length of the DNA molecule exceeds by far the diameter of the nucleus. To fit in the nucleus, DNA is carefully packaged around specific proteins, forming a complex called \emph{chromatin}. Despite a high degree of compaction, DNA, in its uncompressed form, must be rapidly accessible to the protein machineries that regulate the essential functions of life. Recent studies have revealed that chromatin is non-randomly organized within the cell nucleus, and have linked chromatin folding to many vital cellular functions, such as gene regulation, differentiation, DNA replication and genome stability maintenance ~\citep{Dekker2008GeneDimension.,  Therizols2014ChromatinCells., Misteli2004SpatialFunction.}. Hence, understanding the three-dimensional (3D) chromatin conformation is essential for decoding the functions of the genome and can provide a mechanistic explanation of various biological processes and their links to human disease~\citep{Misteli2007BeyondFunction., Mitelman2007TheCausation}. Furthermore, since many biological processes involving DNA are dynamic, there is a need for methodologies that can elucidate the evolution of chromatin conformation over time.


Traditionally, the structure of the genome has been studied using microscopy techniques, such as fluorescent in situ hybridization (FISH)~\citep{vanSteensel2010GenomicsArchitecture.} or, more recently, stochastic optical reconstruction microscopy (STORM,~\citep{Rust2006Sub-diffraction-limitSTORM}) and photoactivated localization microscopy (PALM,~\citep{Gaietta2002MulticolorTrafficking}). Despite the advancements, microscopy approaches are limited to a small number of genomic locations and do not support a comprehensive analysis of the complete genome structure~\citep{Bonev2016}. In recent years, the advancements in chromosome conformation capture (3C)~\citep{Dekker2002CapturingConformation.} have paved the way for the systematic analysis of the 3D structure of chromatin. 3C methods provide measurements of the physical interaction frequencies between fragments of consecutive chromatin loci of a certain resolution, commonly referred to as chromatin bins. \cite{Lieberman-Aiden2009ComprehensiveGenome} proposed \emph{HiC}, a higher-throughput, higher-resolution, 3C-based method that quantifies intra- and inter-chromosomal interaction frequencies at a whole-genome scale. Chromatin interactions captured by HiC are represented as a \emph{contact matrix}, where each entry determines the frequency of interactions between a pair of genomic bins in a population of cells. Therefore, one of the main applications of HiC data is to reconstruct the 3D chromatin structure from the HiC contact matrix.

\paragraph{Related Work} Most methods developed thus far in the literature for this task can be classified into \emph{optimization-based} and \emph{modeling-based} methods. Since, naturally, physical distance and contact frequency are inversely correlated, optimization methods model this relationship with a specific \emph{transfer function} that maps the contact frequencies to distances, yielding a distance matrix. An optimization problem is then formulated to minimize the difference between distances in this matrix and the ones computed by the inferred structures. In practice, this translates into performing multi-dimensional scaling (MDS)~\citep{Kruskal1964MultidimensionalHypothesis} on the distance matrix. Examples of this approach are ~\cite{Duan2010AGenome., Lesne20143DContacts, Wang2012ClassicalScaling, Zhang20133DData, Trieu2014Large-scaleData.b, Rieber2017MiniMDS:Data}.  Modeling-based methods~\citep{Rousseau2011Three-dimensionalSampling, Hu2013BayesianChromosomes, Varoquaux2014AGenome, Zou2016HSA:Structureb, Wang2015InferentialStructure, Park2016ImpactMethods, Oluwadare2018AData.} formulate the relationship between contact frequencies and distances in a probabilistic fashion and perform inference through maximum likelihood estimation or via Bayesian approaches. GEM~\citep{Zhu2018ReconstructingLearning}, a more recent method,   adopts a modified t-SNE algorithm~\citep{VanDerMaaten2008VisualizingT-SNE}  to perform manifold learning.      
        
Nevertheless, several limitations  reduce the usability of the existent methods. First, most methods necessitate the use of a parametric transfer function, which requires making assumptions about the relationship between distances and contact frequencies. One exception is GEM~\citep{Zhu2018ReconstructingLearning}, which, however, fails to preserve the order of the bins in the chromosomes. Second, most methods do not scale with the resolution of recent HiC experiments. High resolution is necessary for accurate, fine-grained structure inference. MiniMDS~\citep{Rieber2017MiniMDS:Data} is the only method designed to address this issue by inferring high-resolution structures for sub-regions of the chromosomes and connecting them together using a low-resolution global structure. Lastly, none of the existing methods can incorporate time-course information and perform \emph{dynamic analysis} of the chromatin structure. In this work we propose a novel manifold learning method, \textbf{REcurrent Autoencoders for CHromatin 3D structure prediction (REACH-3D)}, to infer the dynamic 3D chromatin structure from HiC data.

\section{Methods}
REACH-3D addresses the major challenges that come with HiC data and the limitations of existing methods. Our solution exploits manifold learning as a means to reduce the dimensionality of HiC data and infer the 3D chromatin structure. 
To apply manifold learning to the problem at hand, our method first assumes  that the 3D coordinates of the chromatin bins lie on an embedded, non-linear manifold. The manifold lives in a high-dimensional space, represented by a contact matrix \(\mathbf{X} \in \R^{N \times N}\) through the HiC experiment. Our goal is to map the HiC data \(\mathbf{X}\) to \emph{3D Euclidean coordinates} \(\mathbf{Z} \in \R^{N \times 3} \), corresponding to the intrinsic dimensionality of the HiC data, i.e., the coordinates of the chromatin bins.



\begin{figure}[t!]
    \includegraphics[width=1\textwidth]{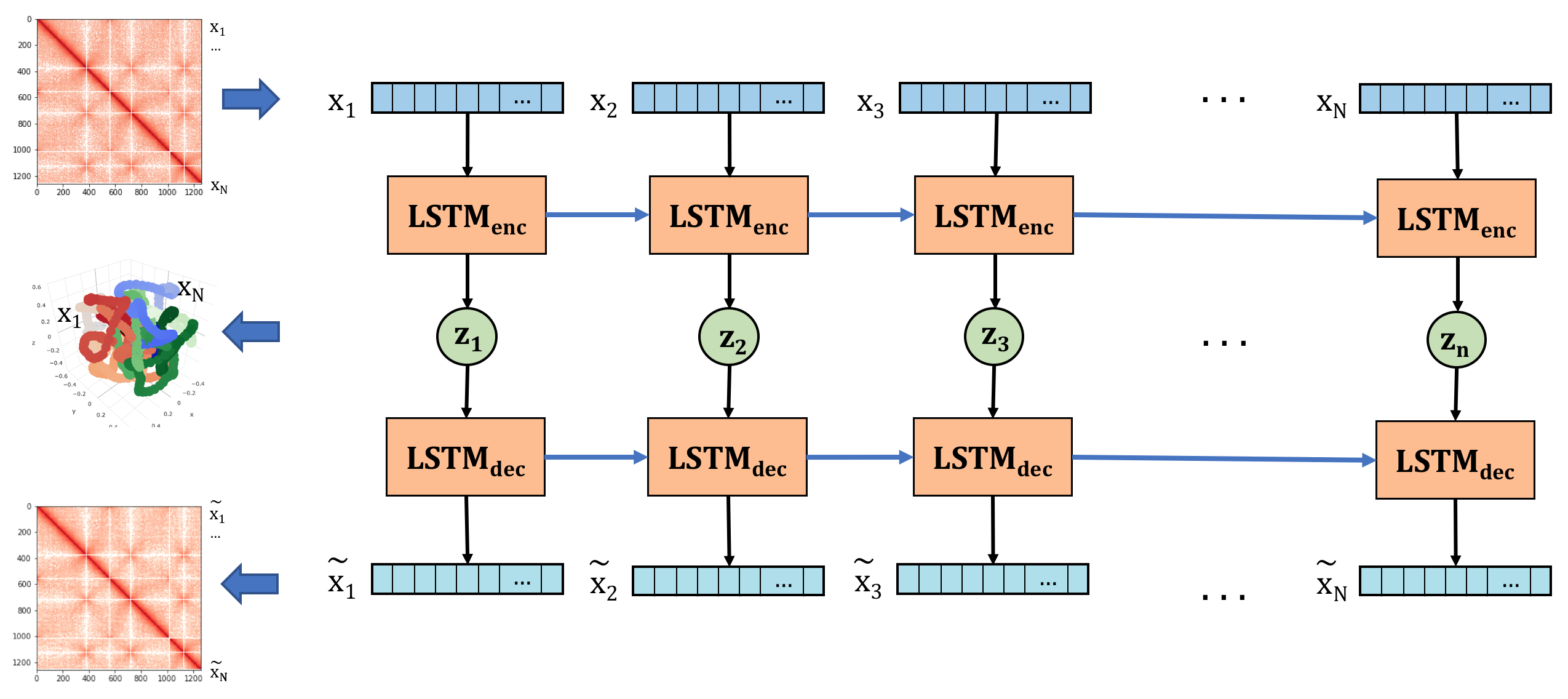}
    \caption{\small{REACH-3D autoencoder architecture with recurrent neural units, specifically LSTMs. The inputs to the encoder cells are the features of each genomic bin. The embedding layer represents the three-dimensional chromatin structure. The outputs of the decoder cells are the reconstructions of the genomic bin features.
    }}
    \label{fig:GRAE_architecture}
\end{figure}

The architecture of REACH-3D is inspired by the sequence-to-sequence models introduced by \cite{Sutskever2014SequenceNetworks}, frequently used in natural language processing for  translation or sentence completion. 
In comparison to a sequence-to-sequence architecture, we encode each element, i.e., each bin in the genomic sequence, into a fixed 3D vector, which is in turn decoded to reconstruct the original element. 
To encode the whole chromatin sequence, REACH-3D consists of a sequence of autoencoders, where each autoencoder is matched to one chromatin bin, thus ensuring that the genomic order of the bins is preserved. As illustrated in Figure \ref{fig:GRAE_architecture}, REACH-3D is designed as a network with recurrent units, specifically the commonly employed Long Short-Term Memory (LSTM) \citep{hochreiter1997long} units; the autoencoders are connected and pass information onward as the sequence progresses. The input to each encoder cell is the feature vector of each bin, representing the interactions of that bin with all other bins in the genome. The sequence length and number of features are equal in this case, and given by \(N\), the number of bins in the chromatin structure. 
            
\paragraph{Encoder}
For the encoder we use an LSTM neural network. The input to each encoder LSTM cell \(i\) is the feature vector of the corresponding element \(i\), \( \mathbf{x_i} = (x_i^1, x_i^2,...,x_i^N) \in \R^N \), where \( x_{i}^{j} \) is the contact frequency between bins \(i\) and \(j\), and the hidden state of the previous encoder cell is \(\mathbf{h_{i-1}^{enc}} \in \R^3 \). The output of the encoder cell, when applying the encoding \(f^{enc}\), is the fixed low-dimensional embedding \( \mathbf{z_i} \in \R^3 \): 
    \begin{equation}
             \mathbf{z_i} = f^{enc}(\mathbf{x_i}, \mathbf{h_{i-1}^{enc}}) 
    \end{equation}
             
\paragraph{Decoder}
Similarly, the decoder is also an LSTM neural network. The input to the decoder LSTM cell is the embedding of the corresponding element \(i\), \( \mathbf{z_i} \in \R^3 \), and the hidden state of the previous decoder cell, \( \mathbf{h_{i-1}^{dec}} \in \R^N \). The output of the decoder cell, when applying the decoding \(g^{enc}\), is the fixed reconstruction of the contact frequencies \( \mathbf{\tilde{x}_i} \in \R^N\):  
    \begin{equation}
        \mathbf{\Tilde{x}_i} = g^{dec}(\mathbf{z_i}, \mathbf{h_{i-1}^{dec}}) = g^{dec}(f^{enc}(\mathbf{x_i}, \mathbf{h_{i-1}^{enc}}),  \mathbf{h_{i-1}^{dec}})
    \end{equation}
   
The obtained sequence of embeddings in 3D space, \( \mathbf{Z} = (\mathbf{z_1}, \mathbf{z_2},..., \mathbf{z_N})\), where \(\mathbf{z_i} \in \R^3\), represents the coordinates of the bins in the predicted chromatin structure, as illustrated in Figure \ref{fig:GRAE_architecture}.

\paragraph{Loss Function}
The loss function in Equation (\ref{equation:vanilla_loss}) is composed of two terms, a main reconstruction loss, \(\mathcal{L}_{rec}\), and a distance loss, \(\mathcal{L}_{d} \).
The reconstruction loss is defined as the standard mean squared error between the input and the reconstruction of the HiC matrix (Equation (\ref{equation:reconstruction_loss})).
The distance loss, inspired by \emph{biological priors}, acts as a regularizer on the lower- and upper-bound of the Euclidean distance between two consecutive bins and safeguards against unreasonably high or low distances. Furthermore, the loss formulation in Equation (\ref{equation:vanilla_loss}) is similar to a Lagrangian expression, and $\lambda$ can be seen as a Lagrange multiplier. 
Specifically, to model the folding behavior of the chromosomes, we introduce two bounds: a lower bound, defined by \(b_{min}\), which represents a fully-packed folding, and an upper bound, defined by \(b_{max}\), which represents a fully-extended folding of the chromosome bins. Hence, Equation (\ref{equation:distance_loss}) can be interpreted as the result of two forces (the deviation from the lower and the upper bound) pulling in opposite directions. This does not imply that the distances are at equal deviation from both lower and upper bounds, since the reconstruction cost can impose a preference towards one of the bounds. 
The lower and upper bound values defining the packing ratio across the bins are independent of the resolution. Changes in the resolution are equivalent to scaling the whole structure, thus obtaining the same effect as proportionally changing the bounds. 

    \begin{equation}
        \mathcal{L} = \mathcal{L}_{rec}(\mathbf{X}, \mathbf{\tilde{X}}) + \lambda \mathcal{L}_{d}(\mathbf{Z})
        \label{equation:vanilla_loss}               
    \end{equation}
    \begin{equation}
        \mathcal{L}_{rec}(\mathbf{X}, \mathbf{\tilde{X}}) = \frac{1}{N} \sum_{i=1}^{N}{\lVert \mathbf{x_i} - \mathbf{\tilde{x}_i}\rVert_{2}^{2}} = \frac{1}{N} \sum_{i=1}^{N}{\lVert\mathbf{x_i} - g^{dec}(f^{enc}(\mathbf{x_i}, h_{i-1}^{enc}),  h_{i-1}^{dec})\rVert_{2}^{2}}
        \label{equation:reconstruction_loss}
    \end{equation}
    \begin{equation}
         \mathcal{L}_{d}(\mathbf{Z}) = \frac{1}{N - 1} \sum_{i=1}^{N - 1}{\left(\max(b_{min} - \lVert \mathbf{z_i}, \mathbf{z_{i+1}}\rVert_{2}^2, 0) + \max(\lVert\mathbf{z_i}, \mathbf{z_{i+1}}\rVert_{2}^2 - b_{max}, 0)\right)}
         \label{equation:distance_loss}
    \end{equation}


In our model, the \(\lambda\) multiplier is a hyperparameter that we have to optimize. The value of \(\lambda\) is evaluated on a few factors: FISH distances validation, loss values, and simulated structures accuracy. All the steps of the inference process are presented in Algorithm \ref{algorithm:GRAE}.
            
            \begin{algorithm}[t!]
                \caption{Inferring chromatin structures with REACH-3D}\label{algorithm:GRAE}
                \small{
                \begin{algorithmic}[1]
                \REQUIRE { \(\textbf{X}\in R^{N \times N}\): HiC contact frequency matrix \\
                \(N\): sequence length and the number of features \\
                \(d\): dimensionality of the embedding \\
                \(E\): number of epochs \\}
                \STATE {Initialize the network weights \(\mathbf{W^{enc}} \), \(\mathbf{W^{dec}}  \) ;}
                \STATE {\(\mathbf{Z} \in R^{N \times d} \) embeddings; Zero state embedding \(\mathbf{z_{0}} = \mathbf{0}\) }
                \FOR {epoch = 1,2, ... , E}
                     \FOR {i = 1,2, ... , N}
                    \STATE {Encode the input \(\mathbf{x_{i}} \in R^{N}\) into the embedding \(\mathbf{z_{i}} \in R^{d}\) }
                    \STATE {Decode \(\mathbf{z_{i}} \in R^{d}\) into the reconstruction \(\mathbf{\tilde{x}_{i}} \in R^{N}\) }
                    
                     \ENDFOR
                     \STATE {Update \(\mathbf{W^{enc}}\), \(\mathbf{W^{dec}}\) using \( \nabla \mathcal{L}\), the gradient of the loss in Equation  (\ref{equation:vanilla_loss}) \\ 
                            }
                \ENDFOR
                \RETURN {\(\mathbf{Z} \in R^{N\times d}\)}
                \end{algorithmic}}
               
            \end{algorithm}
        

\paragraph{Comparison to Related Work} 
REACH-3D is fundamentally different from all existing methods, and shares a few common concepts with GEM~\citep{Zhu2018ReconstructingLearning}. The model is specifically tailored for the problem of 3D chromatin prediction, and directly addresses several limitations of prior approaches, as summarized below:
\begin{itemize}
    \vspace{-0.8em}
    \setlength\itemsep{-0.05em}
    \item  \textbf{Transfer function assumption:} 
    We perform the inference directly on the contact frequency matrix, without requiring any assumptions on the mapping between distances and contacts. 
    
    \item \textbf{Scale with the resolution:} In our model, we employ shared weights in the network architecture, leading to a significantly smaller number of trainable parameters and a shorter network training time.
   
    
    \item \textbf{Sparse contact matrices:} Since dimensionality reduction results in the minimum number of dimensions that best describe the data, sparsity does not affect the structure, provided that the initial dimensions accurately explain the variations in the data. 
    
    \item \textbf{Sequence preservation:} We introduce recurrent neural units in our model to describe the sequential relationship between the elements in our structure, i.e., the bins in the genome.
    
    \item \textbf{Simulating folding mechanism:} Lower and upper bounds of the distance between consecutive bins prohibit non-realistic solutions in terms of folding, and can be parameterized to allow REACH-3D flexibility with respect to different organisms.

    \item \textbf{Global structure recovery:} The autoencoder we propose is able to learn both the local structure through its high expressiveness, but also to preserve the global structure through the memory of the LSTM cells. 

\end{itemize}
     
Since each HiC matrix represents measurements over a \textit{population of cells}, we adopt an \textbf{ensemble prediction} strategy, where we predict a set of potential structures representative of that population. To achieve this, we simply use different weight initializations and repeat the learning process. 
    
\paragraph{Time-dependent Analysis}
HiC data is a \emph{snapshot} of the interactions between and within the chromosomes. Nevertheless, the chromosome folding process is \emph{dynamic} and depends on other biological processes, such as the cell cycle or the differentiation state. Although the focus of the HiC community is shifting towards acquiring data and establishing protocols for 4D analysis~\citep{Dekker2017TheProject}, existing methods are limited to 3D analysis. Applying them to 4D data would be equivalent to obtaining independent 3D structures for each of the time points, thus disregarding the continuous nature of the data. One of the main novelties of our model is that it incorporates information about the structure at previous time points in the inference process. The simplest way of achieving this is via \emph{weight sharing}: for each time point, we initialize the weights of the network with the corresponding values of the previous time point. In this way, we directly model the evolution of the chromatin conformation over time.

\section{Experiments and Results}
The inference of 3D chromatin structure falls under the unsupervised learning paradigm. In contrast to protein folding, which can be assessed using X-ray crystallography or spectroscopy-based techniques, no method able to experimentally determine the folding of the chromatin exists. Therefore, the lack of ground truth requires alternative methods for evaluating the inferred 3D structures. We first tune the hyperparameter \(\lambda\) by looking at the value of the loss (and implicitly, at the reconstruction error) and the presence of the desired biological and physical properties in the structure. To evaluate REACH-3D and compare it with the state-of-the-art methods, we use synthetic data and, where available, 3D FISH microscopy-measured pairwise distances.

\paragraph{Datasets}

There exist numerous experimental HiC datasets on different organisms, cell types and resolutions in the literature. After reviewing existing datasets, we have selected the following: 
        
\begin{enumerate}
    \vspace{-0.8em}
    \setlength\itemsep{-0.05em}
  \item \emph{Synthetic Data:} The only synthetic contact frequency matrix in the literature was developed by \cite{Trussart2015AssessingDomains}. 
  To create a synthetic HiC contact matrix, the authors created 100 3D toy models of a single, hypothetical chromosome of 1 Mb length and aggregated them to derive one single contact frequency matrix, representative of the population. 
    
   \item \emph{Fission Yeast:} A second dataset used for evaluation and performance of REACH-3D comes from fission yeast, and includes HiC measurements at different time points of the fission yeast's cell cycle~\citep{Tanizawa2017ArchitecturalCycle}.
    
  \item \emph{Human:} Last, we experimented on HiC maps of all human chromosomes from a lymphoblastoid cell line (GM12878)~\citep{Rao2014ALooping}, chosen due to its high-resolution data.
\end{enumerate}

\paragraph{Experimental Setup}
REACH-3D is implemented in Tensorflow 1.9~\citep{tensorflow2015-whitepaper}. We use \cite{Kingma2014Adam:Optimization}'s Adam optimizer with the following settings: a learning rate of \(0.001\), an exponential decay rate for the first and second moment estimates of \(0.9\) and \(0.999\) respectively, and an epsilon value of \(1e^{-08}\). We also apply gradient clipping with a clipping value of \(5\). To initialize the weights of the network we use the Xavier initializer~\citep{glorot2010understanding}. The number of iterations necessary for convergence varies between 1000-5000 epochs, depending on the dataset and resolution. The hyperparameter of the REACH-3D model is the \(\lambda \) multiplier. The experiments were ran on an IBM Power System S822LC. 
            
\begin{figure}[t!]
\begin{multicols}{6}
  \subfloat[\(\lambda = 10^{-2}\)]{
    \includegraphics[width=0.157\textwidth]{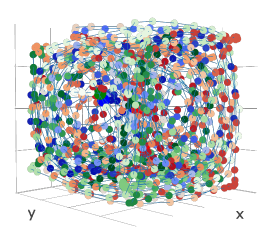}
    }
    \subfloat[\(\lambda = 10^{-3}\)]{
    \includegraphics[width=0.157\textwidth]{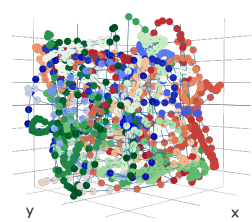}
    }
    \subfloat[\(\lambda = 10^{-4}\)]{
    \includegraphics[width=0.157\textwidth]{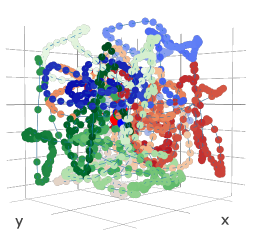}
    }
   \subfloat[\(\lambda = 10^{-5}\)]{
    \includegraphics[width=0.157\textwidth]{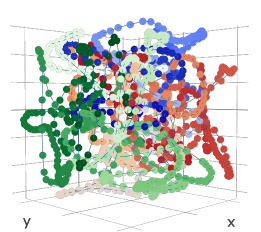}
    }
    \subfloat[\(\lambda = 10^{-6}\)]{
    \includegraphics[width=0.157\textwidth]{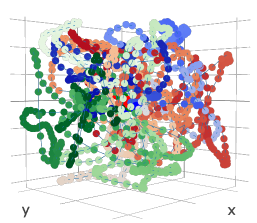}
    }
    \subfloat[\(\lambda = 10^{-7}\)]{
    \includegraphics[width=0.157\textwidth]{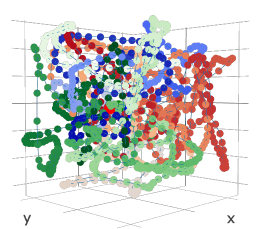}
    }
   
\end{multicols}

\caption{\small{Inferred 3D structures of the fission yeast genome for varying values of the \(\lambda\) multiplier. Each of the three chromosomes is represented by a different color: red, green and blue, respectively. The color gradient represents the sequence of the bins on the chromosomes (light: start, dark: end of the chromosome).}}
\label{fig:Structures_vary_lambda_GSE93198_Tanizawa_10000_20min}
\end{figure}

\paragraph{Hyperparameter Tuning}             
To assess the inferred structures, we initially visually examine biological properties expected from prior knowledge, namely the preservation of genomic bin sequence, the existence of chromosome territories, i.e., subnuclear compartments where each chromosome is localized, and the presence of intra- and inter-chromosomal interactions and long-range loops. Resulting structures for the whole fission yeast genome and various values of \(\lambda\) are shown in Figure \ref{fig:Structures_vary_lambda_GSE93198_Tanizawa_10000_20min}. Larger values of \(\lambda\), e.g. \(10^{-2}\) and \(10^{-3}\), do not fully preserve the genomic bin sequence and the chromosome territories are not clearly visible. As \(\lambda\) decreases, the structures are more consistent with the biological prior expectations. At the same time, the lowest total loss and thus the best reconstruction is obtained for \(\lambda =  10^{-6}\).
          
\subsection{Synthetic Data Validation}    
To evaluate the method, we generate an ensemble of 100 3D structures and compare them to the 100 ground-truth ones. Since there is no one-to-one correspondence between the structures in the ensembles, we use a probabilistic approach, as described in the following. The Euclidean distance between bins \(i\) and \(j\) is denoted as \(d_{ij}\) and \(d^{*}_{ij}\) for ground-truth and predicted structures respectively. Similarly, \(p_{d_{ij}}\), \(p_{d^{*}_{ij}}\) denote the probability distribution of \(d_{ij}\) and \(d^{*}_{ij}\) respectively. 
We first compute \(d_{ij}\) between all bins in all 100 ground-truth structures, i.e., \( d^{s}_{ij} = \lVert z_{i}^{s}, z_{j}^{s} \rVert\), for $s \in \{1, ..., 100\} $, \(\forall \: i,j \leq N \) and the corresponding distribution for each pair of bins \(d_{ij} \sim p_{d_{ij}}\), \(\forall \: i,j \leq N \). We repeat the same process in the ensemble of predicted structures and compute \(d_{ij}^{*s}\), for \(s \in \{1, ..., 100\} \)  and \(p_{d^{*}_{ij}}\), \(\forall \: i,j \leq N \).   If the predicted structures match the ground-truth, then the distance between \( p_{d_{ij}}, p_{d^{*}_{ij}}\) should be small. To quantify this, we estimate the Wasserstein distance~\citep{vallender1974calculation} between the two distributions,  \( W(p_{d_{ij}}, p_{d^{*}_{ij}})\), \(\forall \: i,j \leq N\). The Wasserstein distance was chosen as an appropriate metric since it does not require that the two distributions have the same support.

\begin{figure}[t!]
        \begin{center}
        \begin{minipage}[]{1\textwidth}
            \begin{multicols}{2}
            \subfloat{
            \includegraphics[width=0.5\textwidth]{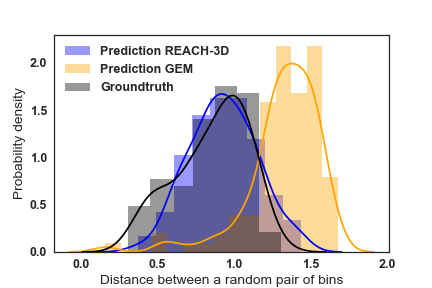}}
            \subfloat{
            \includegraphics[width=0.5\textwidth]{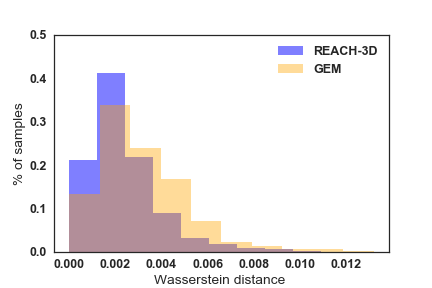}}
        \end{multicols}
        \caption{\small{Evaluation of structures inferred by REACH-3D and GEM for synthetic data at 5 Kb resolution. Left: The distribution of distances between a random pair of bins shows that REACH-3D matches the ground truth, whereas GEM results in a distribution with different shape, mean and variance. Right: The median Wasserstein distance between the distance distributions across all bins are smaller for REACH-3D than GEM.}}
        \label{fig:SyntheticValidation}
        \end{minipage}
        
         \begin{minipage}[]{0.79\textwidth}
         \begin{multicols}{4}
            \subfloat[Ground-truth]{
                    \includegraphics[width=0.23\textwidth]{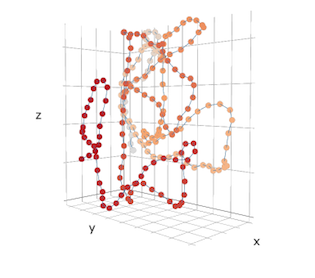}
                }
                \subfloat[REACH-3D]{
                    \includegraphics[width=0.23\textwidth]{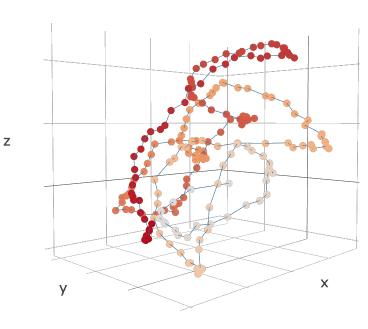}
                }
                \subfloat[miniMDS]{
                    \includegraphics[width=0.26\textwidth]{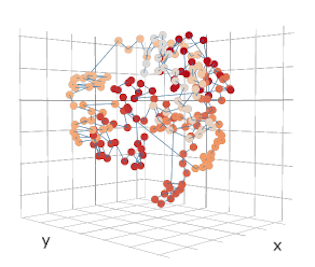}
                }
                \subfloat[GEM]{
                    \includegraphics[width=0.27\textwidth]{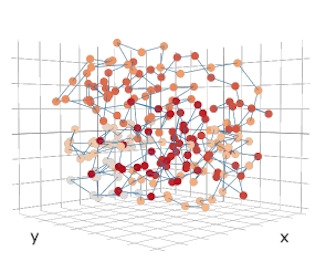}
                }
        \end{multicols}       
        \end{minipage}
            \hfill 
        
            \caption{ \small{One random synthetic structure for a toy chromosome (a) and inferred structures by REACH-3D (b) miniMDS (c) and GEM (d). The color gradient represents the bin sequence (light: start, dark: end).}}
            \label{fig:Simulation_Predictions_5Kb_alpha150}
        \end{center}    
        \end{figure}

We perform the analysis on the synthetic data at 5 Kb resolution and compare the results of REACH-3D using \(\lambda = 10^{-6}\), with the ones from GEM. MiniMDS cannot infer an ensemble of structures and was thus omitted from the comparison. Figure \ref{fig:SyntheticValidation} (a) shows the distribution of the distances for a random pair of bins; REACH-3D obtains a distance distribution very similar to the ground-truth, whereas the distribution obtained by GEM has a markedly different shape, mean and variance. 
Figure \ref{fig:SyntheticValidation} (b) visualizes the distribution of Wasserstein distances across all pairs of bins. The distribution of Wasserstein distances obtained using REACH-3D is right-skewed with a heavy tail, implying that the majority of pairs exhibit a small distance between the ground-truth and predicted distributions, whereas Wasserstein distances computed from GEM are visibly larger. Comparison of the two distributions via the Mann–Whitney U test yielded a p-value of \( 0.0\), indicating that the distributions are indeed statistically significantly unequal. 
        
Finally, we compare the structures obtained by the different algorithms in Figure \ref{fig:Simulation_Predictions_5Kb_alpha150}, and observe that REACH-3D recovers most of the sought properties. Firstly, the sequence of the bins in the chromosome is best illustrated by the REACH-3D structure. MiniMDS recovers only partially the sequence of the bins in the chromosome and GEM results in non-ordered bin positions. In REACH-3D, the long-range interactions are clearly observable, whereas in GEM the bins seem to follow a random behavior and in miniMDS they are recovered only partially. 
        
\subsection{Evaluation on Experimental Data} 
\begin{figure}[t!]
         \begin{minipage}[]{0.45\textwidth}
            \includegraphics[width=0.9\textwidth]{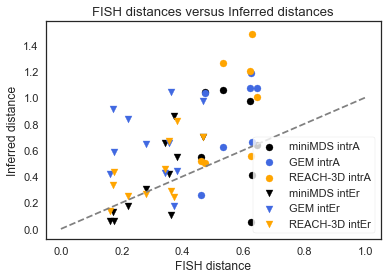}
            \caption{ \small{Scatterplot of FISH and inferred distances between 18 pairs of loci of the fission yeast genome.}}
            \label{fig:GSE93918_FISH_Plot_Correlation}    
        \end{minipage}
            \hfill 
        \begin{minipage}[]{0.45\textwidth}
            \begin{multicols}{2}
                    \subfloat[miniMDS]{
                    \includegraphics[width=0.5\textwidth]{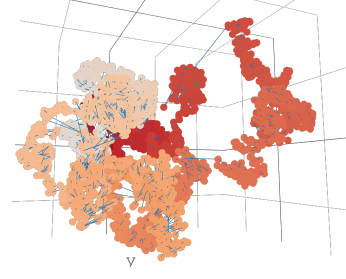}
                    }
                    \subfloat[REACH-3D]{
                    \includegraphics[width=0.5\textwidth]{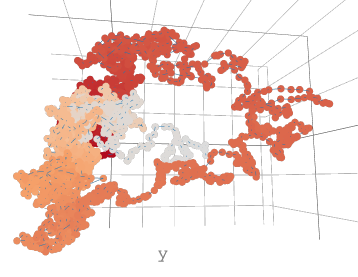}
                    }
                   
                \end{multicols}
                \caption{\small{Inferred 3D structure of human chromosome 22 from a lymphoblastoid cell line by miniMDS (a) and REACH-3D (b).}}
                \label{fig:Predictions_GM12878_miniMDS_paper_10Kb_chr22}
        \end{minipage}

        \end{figure}

\paragraph{Fission Yeast}
In the case of the fission yeast dataset, 3D FISH measurements are available by Tanizawa\cite{Tanizawa2010MappingRegulation}, quantifying pairwise distances in 3D between a limited number of fluorescently-tagged genomic loci. The distances serve as a sparse set of labels of intra- and inter-chromosomal distances, on which the inferred 3D structures can be independently validated. Out of the 18 pairs of loci, 11 are intra- and 7 are inter-chromosomal. We compared the predicted structures of REACH-3D, miniMDS and GEM by computing the Pearson correlation coefficients between FISH distances and Euclidean distances computed from the inferred structures. REACH-3D obtains the highest Pearson correlation coefficient, \(r = 0.76\), followed by miniMDS, \(r = 0.53\), and GEM, \(r = 0.38\). The results are also shown graphically in Figure \ref{fig:GSE93918_FISH_Plot_Correlation}, where we observe similar correlation patterns for both intra- and inter-chromosomal distances. The visual comparison of the inferred structures by REACH-3D, miniMDS and GEM is presented in subsection \ref{subsection:Chromatin_Dynamic_Structure}. 

\paragraph{Human} 
We illustrate the structures of chromosome 22 of the GM12878 cell line at 10 Kb resolution and compare the results of REACH-3D with miniMDS in Figure \ref{fig:Predictions_GM12878_miniMDS_paper_10Kb_chr22}. Due to the high-resolution of the data, GEM was unable to yield results in reasonable computational time. The sequence of the genomic bins is preserved by both methods. Intra-chromosomal interactions, long-range interactions and chromosome looping can be clearly observed in the REACH-3D structure and, to a lesser extent, in the miniMDS structure. 

\subsection{Time-dependent Analysis}\label{subsection:Chromatin_Dynamic_Structure} 
\begin{figure}
      \minipage{1\textwidth}

      \begin{multicols}{8}
        \subfloat[20]{
        \includegraphics[width=0.115\textwidth]{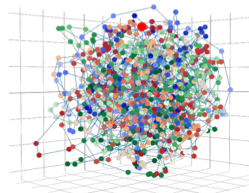}
        }
        \subfloat[30]{
        \includegraphics[width=0.115\textwidth]{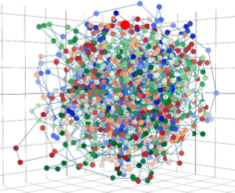}
        }
        \subfloat[40]{
        \includegraphics[width=0.115\textwidth]{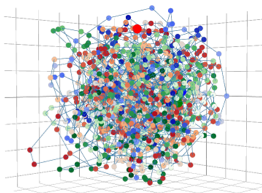}
        }
        \subfloat[50]{
        \includegraphics[width=0.115\textwidth]{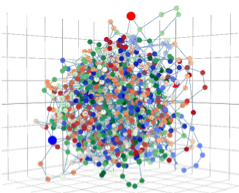}
        }
        \subfloat[60]{
        \includegraphics[width=0.115\textwidth]{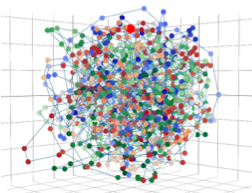}
        }
        \subfloat[70]{
        \includegraphics[width=0.115\textwidth]{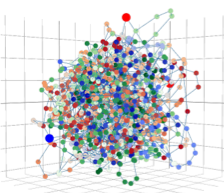}
        }
        \subfloat[80]{
        \includegraphics[width=0.115\textwidth]{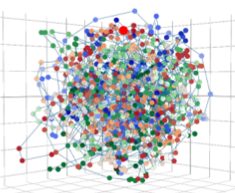}
        }
        \subfloat[120]{
        \includegraphics[width=0.115\textwidth]{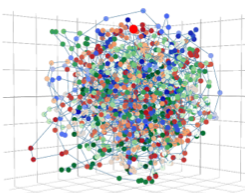}
        }
      \end{multicols}
      
      \endminipage\hfill
      
      \minipage{1\textwidth}

      \begin{multicols}{8}
        \subfloat[20]{
        \includegraphics[width=0.115\textwidth]{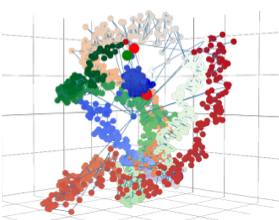}
        }
        \subfloat[30]{
        \includegraphics[width=0.115\textwidth]{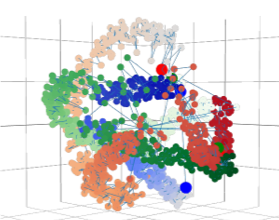}
        }
        \subfloat[40]{
        \includegraphics[width=0.115\textwidth]{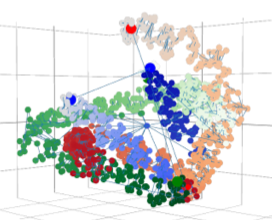}
        }
        \subfloat[50]{
        \includegraphics[width=0.115\textwidth]{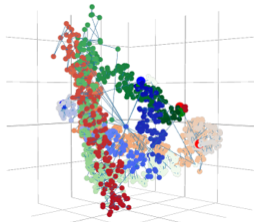}
        }
        \subfloat[60]{
        \includegraphics[width=0.115\textwidth]{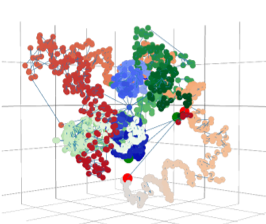}
        }
        \subfloat[70]{
        \includegraphics[width=0.115\textwidth]{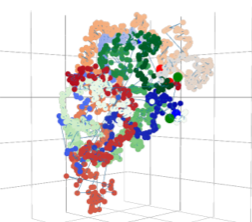}
        }
        \subfloat[80]{
        \includegraphics[width=0.115\textwidth]{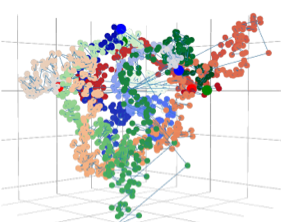}
        }
        \subfloat[120]{
        \includegraphics[width=0.115\textwidth]{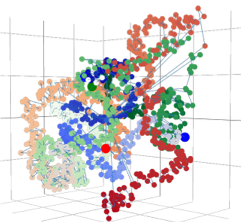}
        }
      \end{multicols}
      
      \endminipage\hfill
      
      \minipage{1\textwidth}

      \begin{multicols}{8}
        \subfloat[20]{
        \includegraphics[width=0.115\textwidth]{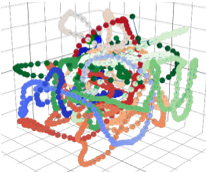}
        }
        \subfloat[30]{
        \includegraphics[width=0.115\textwidth]{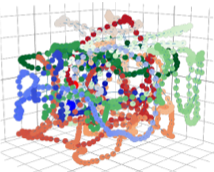}
        }
        \subfloat[40]{
        \includegraphics[width=0.115\textwidth]{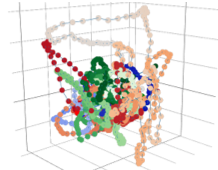}
        }
        \subfloat[50]{
        \includegraphics[width=0.115\textwidth]{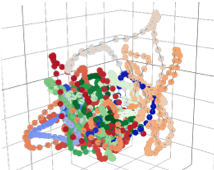}
        }
        \subfloat[60]{
        \includegraphics[width=0.115\textwidth]{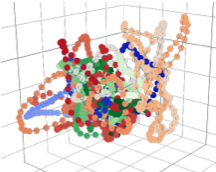}
        }
        \subfloat[70]{
        \includegraphics[width=0.115\textwidth]{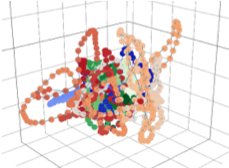}
        }
        \subfloat[80]{
        \includegraphics[width=0.115\textwidth]{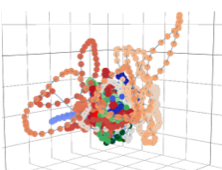}
        }
        \subfloat[120]{
        \includegraphics[width=0.115\textwidth]{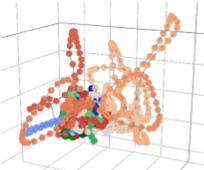}
        }
      \end{multicols}
      
      \endminipage\hfill
      
      \vspace{0.25cm}
      \caption{Inferred 3D chromatin structures for different time points sampled from the fission yeast's cell cycle by GEM (top row), miniMDS (middle row) and REACH-3D (bottom row).}
      \label{fig:Predictions_GSE93198_Tanizawa_10Kb_time}
  \end{figure}

We last perform time series prediction and compare the inferred structures on eight sampled cell cycle time points with miniMDS and GEM, which, as previously mentioned, obtain 3D structures for each time point independently.
The results of \textbf{GEM}  are shown in the first row of Figure \ref{fig:Predictions_GSE93198_Tanizawa_10Kb_time}. The transition between time points cannot be easily assessed, since most of the expected biological and physical properties and, notably, the bin sequence, are not preserved. Therefore, it is hard to distinguish the chromosomes, their interactions and the global 3D chromatin structure. The results of \textbf{miniMDS} are shown in the middle row of Figure \ref{fig:Predictions_GSE93198_Tanizawa_10Kb_time}. MiniMDS preserves the bin sequence in all time points. Nevertheless, the structures are characterized by a zig-zag pattern, and chromosome territories, intra- and inter-chromosomal interactions and long-range loops are not clearly visible in all time points. For example, there are a few structures where the chromosomes are superposed (Figure \ref{fig:Predictions_GSE93198_Tanizawa_10Kb_time} (l) and (o)), and cases where few genomic bins stick out of the structure. More importantly, changes in the structures are rather drastic from one time point to the other, since the prediction of each time point does not exploit the prior information about the already observed structures.
                
The results of \textbf{REACH-3D} with weight sharing are shown in the last row of Figure \ref{fig:Predictions_GSE93198_Tanizawa_10Kb_time}. The structures inferred by REACH-3D best exhibit the expected biological and physical properties, since at all time points the bin sequence is preserved and chromosome territories, intra- and inter-chromosomal interactions and long-range loops are clearly visible. This time, there seems to be a continuum of progression between the structures at different time points, in agreement with gradual changes in chromatin conformation over time. Starting from early M phase, we observe that the structures at 20 and 30 minutes (Figure \ref{fig:Predictions_GSE93198_Tanizawa_10Kb_time} (q) and (r)) are very similar. At late M phase and before the M-to-G1 phase transition (40 minutes, Figure \ref{fig:Predictions_GSE93198_Tanizawa_10Kb_time} (s)), more drastic changes occur, when the left arm of the first chromosome expands and chromosomes two and three are condensed. After M phase and for the remaining of the cell cycle, the structure evolves dynamically. In agreement with observations by \cite{Tanizawa2017ArchitecturalCycle}, the M phase patterns do not drastically disappear, but they rather gradually diminish until the next cell cycle. 
   
       
\section{Conclusions}

In this work, we explore the idea of manifold learning for the inference of the 3D chromatin structure and present a novel method, \emph{REcurrent Autoencoders for CHromatin 3D structure prediction (REACH-3D)}. Our framework addresses the limitations of existing methods by using autoencoders with recurrent neural units to reconstruct the chromatin structure. In comparison to state-of-the-art methods, REACH-3D recovers most faithfully the expected biological properties of the chromatin structure, obtains the highest correlation with microscopy-based distances and the highest reconstruction accuracy on synthetic data. In addition, REACH-3D enables us to perform time series analysis, and thus model the dynamic conformation of chromatin across the cell cycle.
Despite the methodological advancements to infer chromatin structure, new experimental measurements, such as single-cell HiC~\citep{nagano2013single} and microscopy measurements could enable a more direct validation of the results and open the door to other machine learning techniques applicable in the context of supervised learning.
Such models have the potential to advance our understanding of chromatin structure, its various biological functions and their links to human disease.

\bibliographystyle{abbrvnat}
\footnotesize{
\bibliography{references.bib}

\begin{thebibliography}{39}
\providecommand{\natexlab}[1]{#1}
\providecommand{\url}[1]{\texttt{#1}}
\expandafter\ifx\csname urlstyle\endcsname\relax
  \providecommand{\doi}[1]{doi: #1}\else
  \providecommand{\doi}{doi: \begingroup \urlstyle{rm}\Url}\fi

\bibitem[~ et~al.(2010)~, Iwasaki, Tanaka, Capizzi, Wickramasinghe, Lee, Fu,
  and Noma]{Tanizawa2010MappingRegulation}
H.~, O.~Iwasaki, A.~Tanaka, J.~R. Capizzi, P.~Wickramasinghe, M.~Lee, Z.~Fu,
  and K.-i. Noma.
\newblock {Mapping of long-range associations throughout the fission yeast
  genome reveals global genome organization linked to transcriptional
  regulation}.
\newblock \emph{Nucleic Acids Research}, 38\penalty0 (22):\penalty0 8164--8177,
  12 2010.
\newblock ISSN 0305-1048.
\newblock \doi{10.1093/nar/gkq955}.

\bibitem[Abadi et~al.(2015)Abadi, Agarwal, Barham, Brevdo, Chen, Citro,
  Corrado, Davis, Dean, Devin, Ghemawat, Goodfellow, Harp, Irving, Isard, Jia,
  Jozefowicz, Kaiser, Kudlur, Levenberg, Man\'{e}, Monga, Moore, Murray, Olah,
  Schuster, Shlens, Steiner, Sutskever, Talwar, Tucker, Vanhoucke, Vasudevan,
  Vi\'{e}gas, Vinyals, Warden, Wattenberg, Wicke, Yu, and
  Zheng]{tensorflow2015-whitepaper}
M.~Abadi, A.~Agarwal, P.~Barham, E.~Brevdo, Z.~Chen, C.~Citro, G.~S. Corrado,
  A.~Davis, J.~Dean, M.~Devin, S.~Ghemawat, I.~Goodfellow, A.~Harp, G.~Irving,
  M.~Isard, Y.~Jia, R.~Jozefowicz, L.~Kaiser, M.~Kudlur, J.~Levenberg,
  D.~Man\'{e}, R.~Monga, S.~Moore, D.~Murray, C.~Olah, M.~Schuster, J.~Shlens,
  B.~Steiner, I.~Sutskever, K.~Talwar, P.~Tucker, V.~Vanhoucke, V.~Vasudevan,
  F.~Vi\'{e}gas, O.~Vinyals, P.~Warden, M.~Wattenberg, M.~Wicke, Y.~Yu, and
  X.~Zheng.
\newblock {TensorFlow}: Large-scale machine learning on heterogeneous systems,
  2015.
\newblock URL \url{https://www.tensorflow.org/}.
\newblock Software available from tensorflow.org.

\bibitem[Bonev and Cavalli(2016)]{Bonev2016}
B.~Bonev and G.~Cavalli.
\newblock {Organization and function of the 3D genome}.
\newblock \emph{Nature Reviews Genetics}, 17\penalty0 (11):\penalty0 661--678,
  2016.
\newblock ISSN 14710064.
\newblock \doi{10.1038/nrg.2016.112}.

\bibitem[Dekker(2008)]{Dekker2008GeneDimension.}
J.~Dekker.
\newblock {Gene regulation in the third dimension.}
\newblock \emph{Science (New York, N.Y.)}, 319\penalty0 (5871):\penalty0
  1793--4, 3 2008.
\newblock ISSN 1095-9203.
\newblock \doi{10.1126/science.1152850}.

\bibitem[Dekker et~al.(2002)Dekker, Rippe, Dekker, and
  Kleckner]{Dekker2002CapturingConformation.}
J.~Dekker, K.~Rippe, M.~Dekker, and N.~Kleckner.
\newblock {Capturing chromosome conformation.}
\newblock \emph{Science (New York, N.Y.)}, 295\penalty0 (5558):\penalty0
  1306--11, 2 2002.
\newblock ISSN 1095-9203.
\newblock \doi{10.1126/science.1067799}.

\bibitem[Dekker et~al.(2017)Dekker, Belmont, Guttman, Leshyk, Lis, Lomvardas,
  Mirny, O’Shea, Park, Ren, Politz, Shendure, Zhong, and
  Network]{Dekker2017TheProject}
J.~Dekker, A.~S. Belmont, M.~Guttman, V.~O. Leshyk, J.~T. Lis, S.~Lomvardas,
  L.~A. Mirny, C.~C. O’Shea, P.~J. Park, B.~Ren, J.~C.~R. Politz,
  J.~Shendure, S.~Zhong, and t.~D.~N. Network.
\newblock {The 4D nucleome project}.
\newblock \emph{Nature}, 549\penalty0 (7671):\penalty0 219--226, 9 2017.
\newblock ISSN 0028-0836.
\newblock \doi{10.1038/nature23884}.

\bibitem[Duan et~al.(2010)Duan, Andronescu, Schutz, McIlwain, Kim, Lee,
  Shendure, Fields, Blau, and Noble]{Duan2010AGenome.}
Z.~Duan, M.~Andronescu, K.~Schutz, S.~McIlwain, Y.~J. Kim, C.~Lee, J.~Shendure,
  S.~Fields, C.~A. Blau, and W.~S. Noble.
\newblock {A three-dimensional model of the yeast genome.}
\newblock \emph{Nature}, 465\penalty0 (7296):\penalty0 363--7, 5 2010.
\newblock ISSN 1476-4687.
\newblock \doi{10.1038/nature08973}.

\bibitem[Gaietta et~al.(2002)Gaietta, Deerinck, Adams, Bouwer, Tour, Laird,
  Sosinsky, Tsien, and Ellisman]{Gaietta2002MulticolorTrafficking}
G.~Gaietta, T.~J. Deerinck, S.~R. Adams, J.~Bouwer, O.~Tour, D.~W. Laird, G.~E.
  Sosinsky, R.~Y. Tsien, and M.~H. Ellisman.
\newblock {Multicolor and Electron Microscopic Imaging of Connexin
  Trafficking}.
\newblock \emph{Science}, 296\penalty0 (5567):\penalty0 503--507, 4 2002.
\newblock ISSN 00368075.
\newblock \doi{10.1126/science.1068793}.

\bibitem[Glorot and Bengio(2010)]{glorot2010understanding}
X.~Glorot and Y.~Bengio.
\newblock Understanding the difficulty of training deep feedforward neural
  networks.
\newblock In \emph{Proceedings of the thirteenth international conference on
  artificial intelligence and statistics}, pages 249--256, 2010.

\bibitem[Hochreiter and Schmidhuber(1997)]{hochreiter1997long}
S.~Hochreiter and J.~Schmidhuber.
\newblock Long short-term memory.
\newblock \emph{Neural computation}, 9\penalty0 (8):\penalty0 1735--1780, 1997.

\bibitem[Hu et~al.(2013)Hu, Deng, Qin, Dixon, Selvaraj, Fang, Ren, and
  Liu]{Hu2013BayesianChromosomes}
M.~Hu, K.~Deng, Z.~Qin, J.~Dixon, S.~Selvaraj, J.~Fang, B.~Ren, and J.~S. Liu.
\newblock {Bayesian Inference of Spatial Organizations of Chromosomes}.
\newblock \emph{PLoS Computational Biology}, 9\penalty0 (1), 2013.
\newblock ISSN 1553734X.
\newblock \doi{10.1371/journal.pcbi.1002893}.

\bibitem[Kingma and Ba(2014)]{Kingma2014Adam:Optimization}
D.~P. Kingma and J.~Ba.
\newblock {Adam: A Method for Stochastic Optimization}.
\newblock 12 2014.

\bibitem[Kruskal(1964)]{Kruskal1964MultidimensionalHypothesis}
J.~B. Kruskal.
\newblock {Multidimensional scaling by optimizing goodness of fit to a
  nonmetric hypothesis}.
\newblock \emph{Psychometrika}, 29\penalty0 (1):\penalty0 1--27, 3 1964.
\newblock ISSN 0033-3123.
\newblock \doi{10.1007/BF02289565}.

\bibitem[Lesne et~al.(2014)Lesne, Riposo, Roger, Cournac, and
  Mozziconacci]{Lesne20143DContacts}
A.~Lesne, J.~Riposo, P.~Roger, A.~Cournac, and J.~Mozziconacci.
\newblock {3D genome reconstruction from chromosomal contacts}.
\newblock \emph{Nature Methods}, 11\penalty0 (11):\penalty0 1141--1143, 11
  2014.
\newblock ISSN 1548-7091.
\newblock \doi{10.1038/nmeth.3104}.

\bibitem[Lieberman-Aiden and
  Berkum(2009)]{Lieberman-Aiden2009ComprehensiveGenome}
E.~Lieberman-Aiden and N.~v. Berkum.
\newblock {Comprehensive mapping of long range interactions reveals folding
  principles of the human genome}.
\newblock \emph{Science}, 326\penalty0 (5950):\penalty0 289--293, 2009.
\newblock ISSN 1095-9203.
\newblock \doi{10.1126/science.1181369.Comprehensive}.

\bibitem[Misteli(2004)]{Misteli2004SpatialFunction.}
T.~Misteli.
\newblock {Spatial positioning; a new dimension in genome function.}
\newblock \emph{Cell}, 119\penalty0 (2):\penalty0 153--6, 10 2004.
\newblock ISSN 0092-8674.
\newblock \doi{10.1016/j.cell.2004.09.035}.

\bibitem[Misteli(2007)]{Misteli2007BeyondFunction.}
T.~Misteli.
\newblock {Beyond the sequence: cellular organization of genome function.}
\newblock \emph{Cell}, 128\penalty0 (4):\penalty0 787--800, 2 2007.
\newblock ISSN 0092-8674.
\newblock \doi{10.1016/j.cell.2007.01.028}.

\bibitem[Mitelman et~al.(2007)Mitelman, Johansson, and
  Mertens]{Mitelman2007TheCausation}
F.~Mitelman, B.~Johansson, and F.~Mertens.
\newblock {The impact of translocations and gene fusions on cancer causation}.
\newblock \emph{Nature Reviews Cancer}, 7\penalty0 (4):\penalty0 233--245, 4
  2007.
\newblock ISSN 1474-175X.
\newblock \doi{10.1038/nrc2091}.

\bibitem[Nagano et~al.(2013)Nagano, Lubling, Stevens, Schoenfelder, Yaffe,
  Dean, Laue, Tanay, and Fraser]{nagano2013single}
T.~Nagano, Y.~Lubling, T.~J. Stevens, S.~Schoenfelder, E.~Yaffe, W.~Dean, E.~D.
  Laue, A.~Tanay, and P.~Fraser.
\newblock Single-cell hi-c reveals cell-to-cell variability in chromosome
  structure.
\newblock \emph{Nature}, 502\penalty0 (7469):\penalty0 59, 2013.

\bibitem[Oluwadare et~al.(2018)Oluwadare, Zhang, and
  Cheng]{Oluwadare2018AData.}
O.~Oluwadare, Y.~Zhang, and J.~Cheng.
\newblock {A maximum likelihood algorithm for reconstructing 3D structures of
  human chromosomes from chromosomal contact data.}
\newblock \emph{BMC genomics}, 19\penalty0 (1):\penalty0 161, 2018.
\newblock ISSN 1471-2164.
\newblock \doi{10.1186/s12864-018-4546-8}.

\bibitem[Park and Lin(2016)]{Park2016ImpactMethods}
J.~Park and S.~Lin.
\newblock {Impact of data resolution on three-dimensional structure inference
  methods}.
\newblock \emph{BMC Bioinformatics}, 17\penalty0 (1):\penalty0 70, 12 2016.
\newblock ISSN 1471-2105.
\newblock \doi{10.1186/s12859-016-0894-z}.

\bibitem[Rao et~al.(2014)Rao, Huntley, Durand, Stamenova, Bochkov, Robinson,
  Sanborn, Machol, Omer, Lander, and Lieberman~Aiden]{Rao2014ALooping}
S.~S.~P. Rao, M.~H. Huntley, N.~C. Durand, E.~K. Stamenova, I.~D. Bochkov,
  J.~T. Robinson, A.~L. Sanborn, I.~Machol, A.~D. Omer, E.~S. Lander, and
  E.~Lieberman~Aiden.
\newblock {A 3D Map of the Human Genome at Kilobase Resolution Reveals
  Principles of Chromatin Looping}.
\newblock \emph{Cell}, 159:\penalty0 1665--1680, 2014.
\newblock \doi{10.1016/j.cell.2014.11.021}.

\bibitem[Rieber and Mahony(2017)]{Rieber2017MiniMDS:Data}
L.~Rieber and S.~Mahony.
\newblock {MiniMDS: 3D structural inference from high-resolution Hi-C data}.
\newblock \emph{Bioinformatics}, 33\penalty0 (14):\penalty0 i261--i266, 2017.
\newblock ISSN 14602059.
\newblock \doi{10.1093/bioinformatics/btx271}.

\bibitem[Rousseau et~al.(2011)Rousseau, Fraser, Ferraiuolo, Dostie, and
  Blanchette]{Rousseau2011Three-dimensionalSampling}
M.~Rousseau, J.~Fraser, M.~A. Ferraiuolo, J.~Dostie, and M.~Blanchette.
\newblock {Three-dimensional modeling of chromatin structure from interaction
  frequency data using Markov chain Monte Carlo sampling}.
\newblock \emph{BMC Bioinformatics}, 12\penalty0 (1):\penalty0 414, 10 2011.
\newblock ISSN 1471-2105.
\newblock \doi{10.1186/1471-2105-12-414}.

\bibitem[Rust et~al.(2006)Rust, Bates, and
  Zhuang]{Rust2006Sub-diffraction-limitSTORM}
M.~J. Rust, M.~Bates, and X.~Zhuang.
\newblock {Sub-diffraction-limit imaging by stochastic optical reconstruction
  microscopy (STORM)}.
\newblock \emph{Nature Methods}, 3\penalty0 (10):\penalty0 793--796, 10 2006.
\newblock ISSN 1548-7091.
\newblock \doi{10.1038/nmeth929}.

\bibitem[Sutskever et~al.(2014)Sutskever, Vinyals, and
  Le]{Sutskever2014SequenceNetworks}
I.~Sutskever, O.~Vinyals, and Q.~V. Le.
\newblock {Sequence to Sequence Learning with Neural Networks}.
\newblock 9 2014.

\bibitem[Tanizawa et~al.(2017)Tanizawa, Kim, Iwasaki, and
  Noma]{Tanizawa2017ArchitecturalCycle}
H.~Tanizawa, K.-D. Kim, O.~Iwasaki, and K.-i. Noma.
\newblock {Architectural alterations of the fission yeast genome during the
  cell cycle}.
\newblock \emph{Nature Structural {\&} Molecular Biology}, 24\penalty0
  (11):\penalty0 965--976, 10 2017.
\newblock ISSN 1545-9993.
\newblock \doi{10.1038/nsmb.3482}.

\bibitem[Therizols et~al.(2014)Therizols, Illingworth, Courilleau, Boyle, Wood,
  and Bickmore]{Therizols2014ChromatinCells.}
P.~Therizols, R.~S. Illingworth, C.~Courilleau, S.~Boyle, A.~J. Wood, and W.~A.
  Bickmore.
\newblock {Chromatin decondensation is sufficient to alter nuclear organization
  in embryonic stem cells.}
\newblock \emph{Science (New York, N.Y.)}, 346\penalty0 (6214):\penalty0
  1238--42, 12 2014.
\newblock ISSN 1095-9203.
\newblock \doi{10.1126/science.1259587}.

\bibitem[Trieu and Cheng(2014)]{Trieu2014Large-scaleData.b}
T.~Trieu and J.~Cheng.
\newblock {Large-scale reconstruction of 3D structures of human chromosomes
  from chromosomal contact data.}
\newblock \emph{Nucleic acids research}, 42\penalty0 (7):\penalty0 e52, 4 2014.
\newblock ISSN 1362-4962.
\newblock \doi{10.1093/nar/gkt1411}.

\bibitem[Trussart et~al.(2015)Trussart, Serra, Ba{\`{u}}, Junier, Serrano, and
  Marti-Renom]{Trussart2015AssessingDomains}
M.~Trussart, F.~Serra, D.~Ba{\`{u}}, I.~Junier, L.~Serrano, and M.~A.
  Marti-Renom.
\newblock {Assessing the limits of restraint-based 3D modeling of genomes and
  genomic domains}.
\newblock \emph{Nucleic Acids Research}, 43\penalty0 (7):\penalty0 3465--3477,
  4 2015.
\newblock ISSN 1362-4962.
\newblock \doi{10.1093/nar/gkv221}.

\bibitem[Vallender(1974)]{vallender1974calculation}
S.~Vallender.
\newblock Calculation of the wasserstein distance between probability
  distributions on the line.
\newblock \emph{Theory of Probability \& Its Applications}, 18\penalty0
  (4):\penalty0 784--786, 1974.

\bibitem[Van Der~Maaten and Hinton(2008)]{VanDerMaaten2008VisualizingT-SNE}
L.~Van Der~Maaten and G.~Hinton.
\newblock {Visualizing Data using t-SNE}.
\newblock \emph{Journal of Machine Learning Research}, 9:\penalty0 2579--2605,
  2008.

\bibitem[van Steensel and Dekker(2010)]{vanSteensel2010GenomicsArchitecture.}
B.~van Steensel and J.~Dekker.
\newblock {Genomics tools for unraveling chromosome architecture.}
\newblock \emph{Nature biotechnology}, 28\penalty0 (10):\penalty0 1089--95, 10
  2010.
\newblock ISSN 1546-1696.
\newblock \doi{10.1038/nbt.1680}.

\bibitem[Varoquaux et~al.(2014)Varoquaux, Ay, Noble, and
  Vert]{Varoquaux2014AGenome}
N.~Varoquaux, F.~Ay, W.~S. Noble, and J.~P. Vert.
\newblock {A statistical approach for inferring the 3D structure of the
  genome}.
\newblock \emph{Bioinformatics}, 30\penalty0 (12):\penalty0 i26--i33, 2014.
\newblock ISSN 14602059.
\newblock \doi{10.1093/bioinformatics/btu268}.

\bibitem[Wang(2012)]{Wang2012ClassicalScaling}
J.~Wang.
\newblock {Classical Multidimensional Scaling}.
\newblock In \emph{Geometric Structure of High-Dimensional Data and
  Dimensionality Reduction}, pages 115--129. Springer Berlin Heidelberg,
  Berlin, Heidelberg, 2012.
\newblock \doi{10.1007/978-3-642-27497-8{\_}6}.

\bibitem[Wang et~al.(2015)Wang, Xu, and Zeng]{Wang2015InferentialStructure}
S.~Wang, J.~Xu, and J.~Zeng.
\newblock {Inferential modeling of 3D chromatin structure}.
\newblock \emph{Nucleic acids research}, 43\penalty0 (8):\penalty0 e54, 2015.
\newblock ISSN 13624962.
\newblock \doi{10.1093/nar/gkv100}.

\bibitem[Zhang et~al.(2013)Zhang, Li, Toh, and Sung]{Zhang20133DData}
Z.~Zhang, G.~Li, K.-C. Toh, and W.-K. Sung.
\newblock {3D Chromosome Modeling with Semi-Definite Programming and Hi-C
  Data}.
\newblock \emph{Journal of Computational Biology}, 20\penalty0 (11):\penalty0
  831--846, 2013.
\newblock ISSN 1066-5277.
\newblock \doi{10.1089/cmb.2013.0076}.

\bibitem[Zhu et~al.(2018)Zhu, Deng, Hu, Ma, Zhang, Yang, Peng, Kaplan, and
  Zeng]{Zhu2018ReconstructingLearning}
G.~Zhu, W.~Deng, H.~Hu, R.~Ma, S.~Zhang, J.~Yang, J.~Peng, T.~Kaplan, and
  J.~Zeng.
\newblock {Reconstructing spatial organizations of chromosomes through manifold
  learning}.
\newblock \emph{Nucleic Acids Research}, 2 2018.
\newblock ISSN 0305-1048.
\newblock \doi{10.1093/nar/gky065}.

\bibitem[Zou et~al.(2016)Zou, Zhang, and Ouyang]{Zou2016HSA:Structureb}
C.~Zou, Y.~Zhang, and Z.~Ouyang.
\newblock {HSA: integrating multi-track Hi-C data for genome-scale
  reconstruction of 3D chromatin structure}.
\newblock \emph{Genome Biology}, 17\penalty0 (1):\penalty0 40, 12 2016.
\newblock ISSN 1474-760X.
\newblock \doi{10.1186/s13059-016-0896-1}.

\end{thebibliography}
}
\end{document}